\documentclass[twoside]{article}
\usepackage{qic,epsfig}
\usepackage{mathrsfs}
\usepackage{amsmath}

\begin{document}
\setlength{\textheight}{8.0truein}    

\runninghead{NON-MARKOVIAN DECOHERENCE DYNAMICS OF  $\ldots$}
            {J.-H. An $\ldots$}

\normalsize\textlineskip
\thispagestyle{empty}
\setcounter{page}{1}

\copyrightheading{0}{0}{2008}{000--000}

\vspace*{0.88truein}

\alphfootnote

\fpage{1}

\centerline{\bf
NON-MARKOVIAN DECOHERENCE DYNAMICS OF} \vspace*{0.035truein}
\centerline{\bf  ENTANGLED COHERENT STATES}
\vspace*{0.37truein}\centerline{\footnotesize
Jun-Hong An } \vspace*{0.015truein} \centerline{\footnotesize\it
Department of Physics and Center for Quantum Information Science,}
\baselineskip=10pt \centerline{\footnotesize\it National Cheng Kung
University, Tainan 70101, Taiwan} \centerline{\footnotesize\it
Department of Modern Physics, Lanzhou University,}
\baselineskip=10pt \centerline{\footnotesize\it Lanzhou 730000,
People's Republic of China}\vspace*{10pt}

\centerline{\footnotesize Mang Feng}
\vspace*{0.015truein}\centerline{\footnotesize\it Department of
Physics and Center for Quantum Information Science,}
\baselineskip=10pt \centerline{\footnotesize\it National Cheng Kung
University, Tainan 70101, Taiwan} \centerline{\footnotesize\it State
Key Laboratory of Magnetic Resonance and Atomic and Molecular
Physics,} \baselineskip=10pt \centerline{\footnotesize\it Wuhan
Institute of Physics and Mathematics, Chinese Academy of Sciences,
Wuhan 430071, People's Republic of China}

\vspace*{10pt}

\centerline{\footnotesize Wei-Min Zhang \footnote{Email address:
wzhang@mail.ncku.edu.tw}}
\vspace*{0.015truein}\centerline{\footnotesize\it Department of
Physics and Center for Quantum Information Science,}
\baselineskip=10pt \centerline{\footnotesize\it National Cheng Kung
University, Tainan 70101, Taiwan} \centerline{\footnotesize\it
National Center for Theoretical Science, Tainan 70101, Taiwan}

\vspace*{0.225truein} \publisher{(received date)}{(revised date)}

\vspace*{0.21truein}

\abstracts{
We microscopically model the decoherence dynamics of entangled
coherent states of two optical modes under the influence of vacuum
fluctuation. We derive an exact master equation with time-dependent
coefficients reflecting the memory effect of the environment, by
using the Feynman-Vernon influence functional theory in the
coherent-state representation. Under the Markovian approximation,
our master equation recovers the widely used Lindblad equation in
quantum optics. We then investigate the non-Markovian entanglement
dynamics of the two-mode entangled coherent states under vacuum
fluctuation. Compared with the results in Markovian limit, it shows
that the non-Markovian effect enhances the disentanglement to the
initially entangled coherent state. Our analysis also shows that the
decoherence behaviors of the entangled coherent states depend  on
the symmetrical properties of the entangled coherent states as well
as the couplings between the optical fields and the
environment.}{}{}

\vspace*{10pt}

 \vspace*{3pt} \communicate{to be filled by the Editorial}

\vspace*{1pt}\textlineskip    
\section{Introduction}

Optical fields are widely used in quantum communication since
quantum information is almost invariably transmitted using photons.
Experimental quantum teleportation has been realized using the
discrete two-photon polarization entanglement states
\cite{Bouwmeester97} or the continuous two-mode squeezed
entanglement states \cite{Braunstein98,Furusawa98} as quantum
channels \cite{Wolf07}. Another important type of continuous
variable entanglement states, entangled coherent states
\cite{Sanders92,Gerry97,Agarwal03,Pater03,Armour02,Bose06}, has also
been proposed as a potential quantum channel to teleport unknown
quantum states \cite{van01,Wang01,Jeong01}. In this paper, we shall
investigate the non-Markovian decoherence dynamics of the continuous
variable quantum channel in terms of entangled coherent states.

As well known, a realistic analysis of quantum systems for quantum
information processing must take into account decoherence effect.
There has been an increasing interest in describing various
continuous variable quantum channel under noise
\cite{Jakub04,An05,Rossi06,Adesso,Ban06,Goan07}. Conventional
approaches treat the interaction between the quantum system and its
environment perturbatively, which yield equations of motion such as
Redfield or master equations under the Born-Markov approximation
\cite{Redfield65,Lindblad76,Carmichael93}. Although the
approximation has been widely employed in the field of quantum
optics, where the characteristic time of the environmental
correlation function is short compared with that of the system
investigated \cite{Carmichael93}, its validity is experiencing more
and more challenges in facing new experimental evidences
\cite{blatt07}. Moreover, the Born-Markov approximation is in
general invalid in dealing with most condensed-matter problems, for
example, a quantum system hosted in a nanostructured environment
\cite{John94,Aquino04,Falci05, Budini06,mtlee06}, because possible
large coupling constants and long correlation time scales of the
environment both require a non-perturbative treatment. Thus, how to
develop a general non-perturbative microscopic description of open
quantum systems has attracted much attention recently \cite
{Breuer02,Maniscalco07,Chou07,An07}.

In the present work, we shall focus attention on the influence of
vacuum fluctuation on quantum channels in terms of entangled
coherent states. To this end, we model the system as two optical
modes coupled to a bosonic environment at zero temperature. We shall
then develop non-perturbatively a microscopic description to the
decoherence dynamics of such systems. We have noticed that most of
previous theoretical works to explore the decoherence dynamics of
the optical field system relied on Born and/or Markov approximation
\cite{Jakub04,An05,Rossi06,Adesso,Ban06,Goan07}. To derive
non-perturbatively the decoherence dynamics of an open quantum
system, we will employ the Feynman-Vernon influence functional
theory \cite{Feynman63,Caldeira83,Hu92} in the coherent state path
integral formalism \cite{wzhang90}, which enables us to treat both
of the back-actions from the environment to the system and from the
system to the environment self-consistently. After a careful
evaluation of the coherent state path integrals, we obtain an
operator form of the exact master equation with time-dependent
coefficients describing the full non-Markovian dynamics of the
back-actions between the system and the environment.

We then investigate the non-Markovian decoherence dynamics of the
entangled coherent states \cite{Sanders92} using the exact solution
of the reduced density matrix, where the entanglement is measured by
the concurrence \cite{Wootters98}. The non-Markovian effect is
manifested in the short time peak of the time dependent coefficients
in the master equation, which results in an enhancement of the
disentanglement to the entangled coherent states. Indeed, in a
recently published paper \cite{An07}, we have already used the exact
non-Markovian master equation derived in this paper to study the
decoherence dynamics of another type of the continuous variable
entangled states, i.e. entangled squeezed states, where the
entanglement is determined by the logarithmic negativity
\cite{Werner02} rather than the concurrence since the later is not
applicable to entangled squeezed states.

The paper is organized as follows. In Sec. II, we introduce the
model of two optical modes interacting with a common environment in
the coherent-state representation. In Sec. III, we show the detailed
derivation of the influence functional theory to the model. The
exact master equation is derived in Sec. IV. Sec. V is devoted to
the study of entanglement dynamics and the decoherent properties of
quantum channels in terms of the entangled coherent states. Finally,
a brief summary is made in Sec. VI.

\section{ The Hamiltonian of two optical modes in an environment}

Our system includes two separated optical modes subject to a common
vacuum fluctuation, which is relevant to quantum network and has
been widely investigated \cite{pelli,li}. Since we are interested in
the decoherence of the two optical modes mediated by a vacuum
electromagnetic field after the two-mode entangled coherent state is
prepared \cite{Agarwal03,Pater03}, we can omit the terms regarding
the atoms in \cite{pelli,li}. The Hamiltonian of the whole system is
then given by
\begin{equation}
H=H_{S}+H_{E}+H_{I},  \label{HM}
\end{equation}%
where
\begin{align}
&H_{S}=\hbar \omega _{1}a_{1}^{\dag }a_{1}+\hbar \omega
_{2}a_{2}^{\dag
}a_{2}+\hbar \kappa (a_{1}^{\dag }a_{2}+a_{2}^{\dag }a_{1}), \\
&H_{E}=\sum_{k}\hbar \omega _{k}b_{k}^{\dagger }b_{k}, \\
&H_{I}=\sum_{l,k}\hbar (g_{lk}a_{l}^{\dag }b_{k}+g_{lk}^{\ast
}a_{l}b_{k}^{\dag }) \label{HM1},
\end{align}
are, respectively, the Hamiltonians of the two optical modes, the
environment (vacuum fluctuation), and the interaction between them.
The operators $a_{l}$ and $a_{l}^{\dag }$ ($l=1,2$) are the
corresponding annihilation and creation operators of the $l$-th
optical mode with frequency $\omega _{l}$. The parameter $\kappa $
is a coherent tunneling rate of photons between the two optical
systems, such as two cavities \cite{Plenio06,Hollenberg06}, which is
proportional to the overlap of the two wave packets of the optical
fields. Such coupled optical array system recently attracts much
attention \cite{Plenio06,Hollenberg06,Angelakis08} for the possible
materialized in a variety of physical systems, for example, fiber
coupled micro-toroidal cavities \cite{Armani03}, arrays of defects
in photonic band gap materials \cite{Song05} and superconducting
qubits coupled through microwave stripline resonators \cite{Aoki06}.

The environment is modeled, as usual, by a set of harmonic
oscillators identifying the vacuum electromagnetic field with the
annihilation and creation operators $b_{k}$ and $b_{k}^{\dag
}(k=1,2,\cdots )$, $g_{lk}$ are the coupling constants between the
optical modes and the environment. In Eq.~(\ref{HM}) we have also
suppressed the polarization of the fields for both the systems and
the environment. Since most quantum optical experiments are made
currently in low temperature and under vacuum condition, the vacuum
fluctuation should be a main source of decoherence. Therefore, we
take the environment to be at zero temperature throughout this
paper.

To apply the influence functional method to an open quantum system,
the first step towards the dynamics of the reduced system is to
compute the forward and backward propagators between certain initial
and final states of the full system by choosing a convenient
representation. In the present work we use the coherent-state
representation \cite{wzhang90}, in which the basis of the Hilbert
space for the environment consists of multi-mode bosonic coherent
states
\begin{equation}
|\mathbf{z}\rangle =\prod_{k}|z_{k}\rangle ,~~|z_{k}\rangle =\exp
(z_{k}b_{k}^{\dagger })|0_{k}\rangle ,
\end{equation}%
and that for the two optical modes is the two single-mode bosonic
coherent states
\begin{equation}
|\boldsymbol{\alpha }\rangle =\prod_{l=1}^{2}|\alpha _{l}\rangle
,~|\alpha _{l}\rangle =\exp (\alpha _{l}a_{l}^{\dagger
})|0_{l}\rangle ,
\end{equation}%
where the shortened notations for the complex variables, $\mathbf{z}
=(z_{1},z_{2},\cdots )$ and $\boldsymbol{\alpha }=(\alpha
_{1},\alpha _{2})$ , are introduced.

The coherent states defined above are eigenstates of annihilation
operators,
\begin{equation}
b_{k}|z_{k}\rangle =z_{k}|z_{k}\rangle ,~a_{l}|\alpha _{l}\rangle
=\alpha _{l}|\alpha _{l}\rangle .
\end{equation}%
As these coherent states are over-complete, they obey the resolution
of identity,
\begin{equation}  \label{ri}
\int d\mu (\mathbf{z})|\mathbf{z}\rangle \langle \mathbf{z}|=1,~\int
d\mu \left( \boldsymbol{\alpha }\right) |\boldsymbol{\alpha }\rangle
\langle \boldsymbol{\alpha}|=1,
\end{equation}%
where the integration measures are defined by $d\mu
(\mathbf{z})=\prod_{k} e^{-z^*_{k} z_{k}}\frac{dz^*_k dz_{k}}{2\pi
i}$ and $d\mu \left( \boldsymbol{\alpha }\right)
=\prod_{l}e^{-\alpha^*_{l}\alpha _{l}}\frac{d\alpha^*_l d\alpha
_{l}}{2\pi i}$. As it is shown, the bosonic coherent states we used
here are not normalized, and the normalization factors are moved
into the above integration measures, which corresponds to the
Bargmann representation of the complex space. Moreover, these
coherent states are also nonorthogonal,
\begin{equation}
\langle \mathbf{z}|\mathbf{z}^{\prime }\rangle =\exp
(\sum_{k}z^*_{k}z^{\prime }_{k}),~\langle \boldsymbol{\alpha
}|\boldsymbol{\alpha}^{\prime }\rangle =\exp
(\sum_{l}\alpha^*_{l}\alpha^{\prime }_{l}).
\end{equation}
The use of the coherent-state representation makes the evaluation of
path integrals extremely simple. In the coherent-state
representation, the Hamiltonians of the two optical modes, the
environment (vacuum fluctuation), and the interaction between them
are expressed as
\begin{align}
& H_{S}(\boldsymbol{\bar{\alpha}},\boldsymbol{\alpha })=\hbar
\sum_{l=1}^{2}\omega _{l}\bar{\alpha}_{l}\alpha _{l}+\hbar \kappa
(\bar{\alpha}_{1}\alpha _{2}+\bar{\alpha}_{2}\alpha _{1}) ,
\\ &H_{E}( \mathbf{\bar{z}},\mathbf{z})=\sum_{k}\omega
_{k}\bar{z}_{k}z_{k},  \\
& H_{I}( \boldsymbol{\bar{\alpha}},\boldsymbol{\ \alpha
},\mathbf{\bar{z}},\mathbf{z} )= \sum_{lk}(g_{lk}\bar{\alpha}
_{l}z_{k}+g_{lk}^{\ast }\bar{z}_{k}\alpha _{l}) ,  \label{HC}
\end{align}
where $\bar{\mathbf{z}}$ and $\bar{\boldsymbol{\alpha }}$ denote the
complex conjugate of $\mathbf{z}$ and $\boldsymbol{\alpha }$,
respectively. With the above coherent-state representation, we will
present in the next two sections a detailed derivation of the exact
master equation for the reduced density matrix of the two optical
fields that we have simply outlined in our early work \cite{An07}.

\section{The influence functional theory}

\subsection{The influence functional in coherent-state representation}

We follow the influence functional method of \cite{Anastopoulos} by
expressing the density matrix of the composite system as a
double-path coherent state path integral. After eliminating the
degrees of freedom of the environment, we can incorporate all the
environmental effects on the reduced system in a functional integral
named influence functional \cite{Feynman63}. Then the dynamics of
the reduced system will be governed by an effective action retaining
all the influences from the environment in the influence functional.

The total density matrix of the system plus the environment obeys
the quantum mechanical equation $i\hbar \partial \rho _{\mathrm{tot}
}(t)/\partial t=[H,\rho _{\mathrm{tot}}(t)]$, which yields the
formal solution:
\begin{equation}
\rho _{\mathrm{tot}}\left( t\right) =e^{\frac{-iHt}{\hbar }}\rho
_{\mathrm{tot}}\left( 0\right) e^{\frac{iHt}{\hbar }}.
\end{equation}
Different from the coordinate representation in
\cite{Caldeira83,Hu92}, the coherent-state representation leads to,
\begin{align}
&\langle \boldsymbol{\alpha }_{f},\mathbf{z}_{f}|\rho
_{\mathrm{tot}}\left( t\right) |\boldsymbol{\alpha }_{f}^{\prime
},\mathbf{z}_{f}\rangle \notag \\ &~~~=\int d\mu
(\mathbf{z}_{i})d\mu (\boldsymbol{\alpha }_{i})d\mu (\mathbf{z}
_{i}^{\prime })d\mu (\boldsymbol{\alpha }_{i}^{\prime })
 \langle \boldsymbol{\alpha }_{f},\mathbf{z}_{f};t|\boldsymbol{
\alpha }_{i},\mathbf{z}_{i};0\rangle  \notag \\
&~~~~~~~~~\times \langle \boldsymbol{\ \alpha }_{i},
\mathbf{z}_{i}|\rho _{\mathrm{tot}}(0)|\boldsymbol{\alpha
}_{i}^{\prime }, \mathbf{z}_{i}^{\prime }\rangle   \langle
\boldsymbol{\alpha }_{i}^{\prime },\mathbf{ z}_{i}^{\prime
};0|\boldsymbol{\alpha}_{f}^{\prime },\mathbf{z}_{f};t\rangle ,
\label{tot}
\end{align}%
where the resolutions of identity, Eq.(\ref{ri}), has been used. The
density matrix given by Eq.~(\ref{tot}) describes the behavior of
the two optical modes plus the environment as a whole. As we are
only interested in dynamics of the two optical modes, we will work
with the reduced density matrix by integrating over the
environmental variables. We also assume that the initial density
matrix could be factorized into a direct product of the two-mode
state and the environment state $\rho _{\mathrm{tot}}(0)=\rho
(0)\otimes \rho _{E}(0)$, namely, assuming no correlation between
the environment and the system at $t\leq 0$ \cite{Leggett87}. Then
the reduced density matrix fully describing the dynamics of the two
optical modes is given by
\begin{align}
\rho (\boldsymbol{\bar{\alpha}}_{f},\boldsymbol{\alpha }_{f}^{\prime
};t) &=\int d\mu (\boldsymbol{\alpha }_{i})d\mu (\boldsymbol{\alpha
}_{i}^{\prime})\rho (\boldsymbol{\bar{\alpha}}_{i},
\boldsymbol{\alpha }_{i}^{\prime };0)\notag \\
&~~~~~~~~~\times
\mathcal{J}(\boldsymbol{\bar{\alpha}}_{f},\boldsymbol{\alpha }
_{f}^{\prime };t|\boldsymbol{\bar{\alpha}}_{i},\boldsymbol{\alpha }
_{i}^{\prime };0)  ,  \label{rout}
\end{align}%
where $\rho (\boldsymbol{\bar{\alpha}},\boldsymbol{\alpha }^{\prime
};\tau )\equiv \int d\mu (\mathbf{z})\langle \boldsymbol{\alpha
},\mathbf{z}|\rho _{\mathrm{tot}}(\tau )|\boldsymbol{\alpha
}^{\prime },\mathbf{z}\rangle $, and
\begin{align}
\mathcal{J}(\boldsymbol{\bar{\alpha}}_{f},& \boldsymbol{\alpha
}_{f}^{\prime };t |\boldsymbol{\bar{\alpha}}_{i},\boldsymbol{\alpha
}_{i}^{\prime };0) \notag \\
& =\int d\mu (\mathbf{z}_{f})d\mu (\mathbf{z}_{i})d\mu
(\mathbf{z}_{i}^{\prime }) \langle \boldsymbol{\alpha
}_{f},\mathbf{z}_{f};t|\boldsymbol{\alpha
}_{i},\mathbf{z}_{i};0\rangle \notag  \\ &~~~~~~~~~~~~~\times \rho
_{E}(\mathbf{\bar{z}}_{i},\mathbf{z}_{i}^{\prime };0)\langle
\boldsymbol{\alpha }_{i}^{\prime },\mathbf{z}_{i}^{\prime
};0|\boldsymbol{\alpha }_{f}^{\prime },\mathbf{z}_{f};t\rangle ,
 \label{effp}
\end{align}%
is the propagating function of the reduced density matrix, which
contains two propagators for the total system: the forward and
backward propagators, $ e^{\mp \frac{iHt}{\hbar }}$, plus the
initial density matrix of the environment as a matrix element in the
coherent-state representation.

In the following we will show how  to calculate the forward
propagator in terms of the coherent state path integral
\cite{Klauder79,wzhang90}. The similar calculation could be done for
the backward one. To evaluate the forward propagator operator $e^{
\frac{-iHt}{\hbar }}$ between the initial ($|\boldsymbol{\alpha
}_{i}, \mathbf{z}_{i}\rangle $) and the final ($\langle
\boldsymbol{\alpha }_{f}, \mathbf{z}_{f}|$) coherent states, one can
generally divide the time interval $t_{f}-t_{i}$ into $N$
subintervals. Then by inserting the resolution of identity ($N-1$)
times between each subintervals and taking the limit of large $N$,
we have the forward propagator in terms of the coherent state path
integral,
\begin{align}
\langle \boldsymbol{\alpha }_{f},\mathbf{z}_{f};t|\boldsymbol{\
\alpha }_{i}, \mathbf{z}_{i};0\rangle & = \int
D^{2}\mathbf{z}D^{2}\boldsymbol{\ \alpha } \exp \Big\{\frac{i}{\hbar
} \big(S_{S}[\boldsymbol{\bar{\alpha}},\boldsymbol{\alpha }] \notag
\\ &+S_{I}[\mathbf{\bar{z}},\mathbf{z},\boldsymbol{\bar{\alpha}},
\boldsymbol{\alpha}]+S_{E}[\mathbf{\ \bar{z}},\mathbf{z}]\big)
\Big\},   \label{prop}
\end{align}%
where $S_{S}$, $S_{E}$, and $S_{I}$ are the actions corresponding to
the two optical modes, the environment, and the interaction
Hamiltonian $H_{S}$, $ H_{E}$, and $H_{I}$, respectively,
\begin{align}
S_{S}[\boldsymbol{\bar{\alpha}},\boldsymbol{\alpha
}]=&\sum_{l}\Big\{
-i\hbar \bar{\alpha}_{l}\alpha _{l}\left( t\right)   \notag \\
&~~~~+\int_{0}^{t}d\tau \lbrack i\hbar \bar{\alpha}_{l}\dot{\alpha}
_{l}(\tau )-H_{S}(\boldsymbol{\bar{\alpha}},\boldsymbol{\alpha
})]\Big\},  \label{acts} \\
S_{E}[\mathbf{\bar{z}},\mathbf{z}] =&\sum_{k}\Big\{-i\hbar \bar{z}
_{k}z_{k}(t)  \notag \\
&~~~~+\int_{0}^{t}d\tau \lbrack i\hbar \bar{z}_{k}\dot{z}_{k}(\tau
)-H_{E}(\mathbf{\bar{z}},\mathbf{z})]\Big\},   \label{acte} \\
S_{I}[\mathbf{\bar{z}},\mathbf{z},
\boldsymbol{\bar{\alpha}},&\boldsymbol{ \alpha }]=-\int_{0}^{t}d\tau
H_{I}(\boldsymbol{\bar{\alpha}},\boldsymbol{\ \alpha
},\mathbf{\bar{z}},\mathbf{z}).  \label{actI}
\end{align}%
All the functional integrations are carried out over paths
$\mathbf{\bar{z}} (\tau )$, $\mathbf{z}(\tau )$,
$\boldsymbol{\bar{\alpha}}(\tau )$, and $ \boldsymbol{\alpha }(\tau
)$ with endpoints $\mathbf{\bar{z}}(t)=\mathbf{\ \bar{z}}_{f}$,
$\mathbf{z}(0)=\mathbf{\ z}_{i}$, $\boldsymbol{\bar{\alpha}}
(t)=\boldsymbol{\alpha }_{f}$, and $\boldsymbol{\alpha
}(0)=\boldsymbol{\ \alpha }_{i}$. Substituting Eq.~(\ref{prop}) and
a similar expression for the backward propagator into
Eq.~(\ref{effp}), we obtain
\begin{align}
\mathcal{J}(\boldsymbol{\bar{\alpha}}_{f},\boldsymbol{\alpha
}_{f}^{\prime };t|\boldsymbol{\bar{\alpha}}_{i},& \boldsymbol{\alpha
}_{i}^{\prime };0)=\int D^{2}\boldsymbol{\alpha
}D^{2}\boldsymbol{\alpha }^{\prime }\exp \Big\{\frac{
i}{\hbar }(S_{S}[\boldsymbol{\bar{\alpha}},\boldsymbol{\alpha }]  \notag \\
&-S_{S}^{\ast }[\boldsymbol{\bar{\alpha}}^{\prime },\boldsymbol{ \
\alpha }^{\prime }])\Big\}\mathcal{F[}\boldsymbol{\bar{\alpha}},
\boldsymbol{\ \alpha ,\bar{\alpha}}^{\prime },\boldsymbol{\alpha
}^{\prime }],
\end{align}
where
\begin{align}
&\mathcal{F[}\boldsymbol{\bar{\alpha}},\boldsymbol{\alpha
,\bar{\alpha}} ^{\prime },\boldsymbol{\alpha }^{\prime }]=\int d\mu
(\mathbf{z}_{f})d\mu ( \mathbf{z}_{i})d\mu (\mathbf{z}_{i}^{\prime
})D^{2}\mathbf{z}D^{2}\mathbf{z}^{\prime }  \notag \\
&~~~~~~~\times \rho _{E}(\mathbf{\bar{z}}_{i},\mathbf{z}_{i}^{\prime
};0) \exp \Big\{\frac{i}{\hbar}(S_{E}[\mathbf{\bar{z}},\mathbf{z}]
-S_{E}^{\ast }[\mathbf{\bar{z}}^{\prime },\mathbf{z}^{\prime }]  \notag \\
&~~~~~~~~~~~~~~~+S_{I}[
\mathbf{\bar{z}},\mathbf{z},\boldsymbol{\bar{\alpha}},\boldsymbol{\alpha
}] -S_{I}^{\ast }[\mathbf{\bar{z}}^{\prime },\mathbf{z}^{\prime },
\boldsymbol{\bar{\alpha}}^{\prime },\boldsymbol{\alpha }^{\prime
}])\Big\} \label{influ}
\end{align}%
is defined as the Feynman-Vernon influence functional in the
coherent state representation, which contains all the environmental
effects on the two optical modes.

\subsection{Evaluation of the influence functional}

Now we can calculate explicitly the influence functional of our
model using the coherent-state path-integral formalism presented
above. Substituting the model Hamiltonian into the actions of
Eq.~(\ref{acts}-\ref{actI}), we obtain the explicit form of the
forward propagator. The path integral of the environmental part of
the propagator can be done by the stationary phase method
\cite{Klauder79,wzhang90} with the boundary conditions
$z_{k}(0)=z_{ki}$ and $\bar{z}_{k}(t)=\bar{z}_{kf}$, which results
in the equations of motion,
\begin{equation}
\dot{z}_{k}+i\omega _{k}z_{k}=-i\sum_{l}g_{lk}^{\ast }\alpha
_{l},\text{ \ } \dot{\bar{z}}_{k}-i\omega
_{k}\bar{z}_{k}=i\sum_{l}g_{lk}\bar{\alpha}_{l}, \label{ss}
\end{equation}%
where the paths regarding $\boldsymbol{\bar{\alpha}}$ and
$\boldsymbol{\ \alpha }$ are taken as external sources.  The
solution to the stationary path equation (\ref{ss}) are
\begin{align}
& z_{k}(\tau)=z_{ki}e^{-i\omega_{k}\tau}-i \sum_l g_{l
k}^{*}\int_{0}^{\tau}d\tau'
e^{-i\omega_{k}(\tau-\tau')}\alpha_{l}(\tau'), \label{solfw1}\\
& \bar{z}_{k}(\tau)=\bar{z}_{kf}e^{i\omega_{k} (\tau-t)} -i \sum_l
g_{l k}\int_{\tau}^{t}d\tau'e^{i\omega_{k}
(\tau-\tau')}\bar{\alpha}_{l}(\tau'). \label{solfw2}
\end{align}
Note that the prefactor under the contribution of stationary path in
the coherent-state path integral is unity, and the stationary phase
method to treat the environmental part here is exact for the action
being only a quadratic function of the dynamical variables. The path
integral of the environmental part for the backward propagator
$\langle \boldsymbol{\alpha }_{i}^{\prime },\mathbf{z} _{i}^{\prime
};0|\boldsymbol{\alpha }_{f}^{\prime },\mathbf{z}_{f};t\rangle $ can
be evaluated in the same way.

Since we only consider the vacuum fluctuation, the environment is
initially in the equilibrium state at zero temperature, we then have
$\rho _{E}(\mathbf{\bar{z}}_{i},\mathbf{z} _{i}^{\prime };0)=1$.
Substituting the solution (\ref{solfw1}-\ref{solfw2}) for
$z_{k}(\tau)$, $\bar{z}_{k}(\tau)$ and a similar solution for
$\bar{z}'_{k}(\tau)$, $z'_{k}(\tau)$ together into Eq.~(\ref
{influ}), and using the Gaussian integral identity $\int
\frac{d^{2}z}{\pi } e^{-\gamma \bar{z}z+\lambda z+\nu
\bar{z}}=\frac{1}{\gamma }e^{\frac{\lambda \nu }{\gamma }}$
repeatedly for the integral over ${\bf z}_{i},{\bf z}'_{i}, {\bf
z}_{f}$, we reach the final form of the influence functional that we
have used in \cite{An07},
\begin{align}
&\mathcal{F[}\boldsymbol{\bar{\alpha}},\boldsymbol{\alpha
,\bar{\alpha}} ^{\prime },\boldsymbol{\alpha }^{\prime }]
=  \notag \\
&\exp \Big\{\int_{0}^{t}d\tau \int_{0}^{\tau }d\tau ^{\prime }\Big[
\sum_{l,m}(\bar{\alpha}_{l}^{\prime }(\tau )-\bar{\alpha}_{l}(\tau
))\mu_{lm}(\tau -\tau ^{\prime })  \notag \\
& ~\times \alpha _{m}(\tau ^{\prime })+(\alpha _{l}(\tau )-\alpha
_{l}^{\prime }(\tau ))\mu _{lm}^{\ast }(\tau -\tau ^{\prime
})\bar{\alpha}_{m}^{\prime }(\tau ^{\prime })\Big]\Big\},
\end{align}%
where $\mu _{lm}(x)=\sum_{k}e^{-i\omega _{k}x}g_{lk}g_{mk}^{\ast }$
is the dissipation-noise kernel.

\section{The exact non-Markovian master equation}

\subsection{The propagating function of the reduced density matrix}

In the above derivation of the influence functional, the
back-actions between the two optical modes and the environment have
been treated self-consistently. All the effects from environment on
the two optical modes are incorporated in the influence functional
which leads to a modification to the action of the two optical
modes,
\begin{align}
\mathcal{J}(\boldsymbol{\bar{\alpha}}_{f},\boldsymbol{\alpha
}_{f}^{\prime };t|\boldsymbol{\bar{\alpha}}_{i},\boldsymbol{\alpha
}_{i}^{\prime };0)=\int D^{2}\boldsymbol{\alpha
}D^{2}\boldsymbol{\alpha }^{\prime }\exp \Big\{\sum_{l=1}^{2}
\big(\bar{\alpha}_{l}\alpha _{l}\left( t\right)  \notag \\
+\bar{\alpha}_{l}^{\prime }\alpha _{l}^{\prime }\left( t\right)
\big) -\int_{0}^{t}d\tau
\big[\sum_{l=1}^{2}\big(\bar{\alpha}_{l}\dot{\alpha}_{l}+
\dot{\bar{\alpha}}_{l}^{\prime }\alpha _{l}^{\prime }\big)+iH_{S}(
\boldsymbol{\bar{\alpha}},\boldsymbol{\alpha })  \notag \\
-iH_{S}(\boldsymbol{\bar{\alpha}}^{\prime },\boldsymbol{\alpha }
^{\prime }) \big]\Big\}\mathcal{F[}\boldsymbol{\bar{\alpha}},
\boldsymbol{\ \alpha ,\bar{\alpha}}^{\prime },\boldsymbol{\alpha
}^{\prime }].~~ \label{J}
\end{align}%
To execute the path integral of Eq.~(\ref{J}), again we resort to
the stationary phase method and obtain the equations of motion as
($l\neq l^{\prime }$ )
\begin{align}
\dot{\alpha}_{l}+i(\omega _{l}\alpha _{l}+\kappa \alpha _{l^{\prime
}}) =-\int_{0}^{\tau }d\tau ^{\prime }\sum_{m=1}^{2}\mu _{lm}\left(
\tau -\tau
^{\prime }\right) \alpha _{m}\left( \tau ^{\prime }\right) ,  \label{alpha1} \\
\dot{\bar{\alpha}}_{l}^{\prime }-i(\omega
_{l}\bar{\alpha}_{l}^{\prime }+\kappa \bar{\alpha}_{l^{\prime
}}^{\prime }) =-\int_{0}^{\tau }d\tau ^{\prime }\sum_{m=1}^{2}\mu
_{lm}^{\ast }\left( \tau -\tau ^{\prime }\right)
\bar{\alpha}_{m}^{\prime }\left( \tau ^{\prime }\right) .
\label{alpha2}
\end{align}%
with the boundary conditions $\alpha _{l}\left( 0\right) =\alpha
_{li}$ and $\bar{\alpha}_{l}^{\prime }\left( 0\right)
=\bar{\alpha}_{li}^{\prime }$.

The integro-differential equations render the reduced dynamics
non-Markovian, with the memory of the environment's dynamics
registered in the time-nonlocal kernels. To simplify the discussion,
we further assume that the two optical modes are identical, i.e.,
$\omega _{1}=\omega _{2}\equiv \omega _{0}$. Then the coupling
strength to the common environment should also be the same:
$g_{1k}=e^{i\phi }g_{2k}\equiv g_{k}$, where the phase factor
$e^{i\phi }\equiv \lambda $ models the phase difference between the
two optical modes coupling with the environment. In the present work
we will consider two special cases: the two optical modes couple
with the environment in phase (a constructive interference coupling
with $\phi =0\rightarrow \lambda =1$) and out of phase (a
destructive interference coupling with $\phi =\pi $ so that $\lambda
=-1$). By introducing the new variables
\begin{align}
\alpha _{l}( \tau ) =\alpha _{li}u( \tau) -\alpha _{l'i}v ( \tau )
,\notag \\ \bar{\alpha}'_{l}( \tau ) =\bar{\alpha}'_{li}
\bar{u}(\tau) -\bar{\alpha}'_{l'i}\bar{v}(\tau) , \label{soluen}
\end{align}
and using the equations of motion (\ref{alpha1}-\ref{alpha2}), we
obtain the propagating function of the reduced density matrix as
\begin{align}
\mathcal{J}(& \boldsymbol{\bar{\alpha}}_{f},\boldsymbol{\alpha
}_{f}^{\prime };t|\boldsymbol{\bar{\alpha}}_{i},\boldsymbol{\alpha
}_{i}^{\prime };0)= \notag
\\
& \exp \Big\{\sum_{l=1}^{2}\big[u\bar{\alpha}_{lf}\alpha _{li}
+\bar{u}\bar{\alpha}_{li}^{\prime }\alpha _{lf}^{\prime
}-(\bar{u}u+\bar{v}v-1)\bar{\alpha}_{li}^{\prime }\alpha _{li}\big]
\notag
\\
&~~~~~-\sum_{\langle l,l^{\prime }\rangle }^{2}\big[v\bar{\alpha}
_{lf}\alpha _{l^{\prime }i}+\bar{v}\bar{\alpha}_{li}^{\prime }\alpha
_{l^{\prime }f}^{\prime }-(\bar{u}v
+\bar{v}u)\bar{\alpha}_{li}^{\prime }\alpha_{l^{\prime
}i}\big]\Big\}.  \label{prord}
\end{align}%
where $u, v$ are solutions of Eq.~(\ref{soluen}) at time $\tau=t$.
The exact reduced density matrix is then easy to be obtained by
substituting the above solution of the propagating function into
Eq.~(\ref{rout}) and integrating over the initial state.

\subsection{The exact non-Markovian master equation}

Eq.~(\ref{prord}) is an exact result. In this section, we will
deduce the master equation from Eqs.~(\ref{rout}) and (\ref{prord}).
From Eq.(\ref{prord}), 
we obtain
\begin{equation}
\alpha _{li}\mathcal{J}=\frac{u\frac{\delta \mathcal{J}}{\delta
\bar{\alpha} _{lf}}+v\frac{\delta \mathcal{J}}{\delta
\bar{\alpha}_{l^{\prime }f}}}{ u^{2}-v^{2}},\text{ \
}\bar{\alpha}_{li}^{\prime }\mathcal{J}=\frac{\bar{u} \frac{\delta
\mathcal{J}}{\delta \alpha _{lf}^{\prime }}+\bar{v}\frac{\delta
\mathcal{J}}{\delta \alpha _{l^{\prime }f}^{\prime
}}}{\bar{u}^{2}-\bar{v} ^{2}},  \label{fi}
\end{equation}%
which will be used to eliminate the dependence on the initial values
$\boldsymbol{\bar{\alpha}}_{i},\boldsymbol{\alpha }_{i}^{\prime }$
in Eq.~(\ref{rout}). Combining Eqs.~(\ref{prord}) and (\ref{rout})
together, and using the identities of Eq.~(\ref{fi}), the evolution
equation of the reduced density matrix is given by
\begin{align}
&&\dot{\rho}(\boldsymbol{\bar{\alpha}},\boldsymbol{\alpha }^{\prime
};t) =\sum_{l=1}^{2}\Big\{-i\Omega
(t)\big[\bar{\alpha}_{l}\frac{\delta \rho (
\boldsymbol{\bar{\alpha}},\boldsymbol{\alpha }^{\prime };t)}{\delta
\bar{\alpha}_{l}}-\frac{\delta \rho
(\boldsymbol{\bar{\alpha}},\boldsymbol{\alpha
}^{\prime };t)}{\delta \alpha _{l}}\alpha _{l}\big]  \notag  \label{meics} \\
&&+\Gamma (t)\big[2\frac{\delta ^{2}\rho (\boldsymbol{\bar{\alpha}},
\boldsymbol{\ \alpha }^{\prime };t)}{\delta \alpha _{l}\delta
\bar{\alpha} _{l}}-\bar{\alpha}_{l}\frac{\delta \rho
(\boldsymbol{\bar{\alpha}}, \boldsymbol{\alpha }^{\prime
};t)}{\delta \bar{\alpha}_{l}}-\frac{\delta \rho
(\boldsymbol{\bar{\alpha}},\boldsymbol{\alpha }^{\prime };t)}{\delta
\alpha _{l}}\alpha _{l}\big]\Big\}  \notag \\
&&+\sum_{\langle l,l^{\prime }\rangle }^{2}\Big\{-i\Omega ^{\prime
}(t)\big[ \bar{\alpha}_{l}\frac{\delta \rho
(\boldsymbol{\bar{\alpha}},\boldsymbol{ \alpha }^{\prime
};t)}{\delta \bar{\alpha}_{l^{\prime }}}-\frac{\delta \rho (
\boldsymbol{\bar{\alpha}},\boldsymbol{\alpha }^{\prime };t)}{\delta
\alpha_{l}}\alpha _{l^{\prime }}\big]  \notag \\
&&+\Gamma ^{\prime }(t)\big[2\frac{\delta ^{2}\rho
(\boldsymbol{\bar{\alpha}} ,\boldsymbol{\alpha }^{\prime
};t)}{\delta \bar{\alpha}_{l}\delta \alpha _{l^{\prime
}}}-\bar{\alpha}_{l}\frac{\delta \rho (\boldsymbol{\bar{\alpha}},
\boldsymbol{\alpha }^{\prime };t)}{\delta \bar{\alpha}_{l^{\prime
}}}-\frac{ \delta \rho (\boldsymbol{\bar{\alpha}},\boldsymbol{\alpha
}^{\prime };t)}{\delta \alpha _{l}}\alpha _{l^{\prime }}\big]\Big\}  \notag \\
&&
\end{align}%
where
\begin{align}
& \Gamma (t)+i\Omega (t) = -
\frac{u\dot{u}-v\dot{v}}{u^{2}-v^{2}}, \nonumber \\
& \Gamma'(t)+i\Omega'(t)= - \frac{v \dot{u}-u\dot{v}}{u^{2}-v^{2}}.
\label{go}
\end{align}
Eq.~(\ref{meics}) is the exact master equation of the reduced
density matrix for the dynamics of the two optical modes in the
coherent-state representation, in which $\Omega (t)$ plays the role
of a shifted time-dependent frequency of the two modes, $\Omega
^{\prime }(t)$ accounts for a shifted time-dependent coherent
interaction between the two modes, $ \Gamma (t)$ represents a
time-dependent individual decay rate of each mode, and $\Gamma
^{\prime }(t)$ is for a correlated decay rate of the two modes
induced by the environment.

If we define a new variable $F_{\pm }(\tau )=u(\tau )\pm v(\tau )$,
then Eqs.~(\ref{alpha1}-\ref{alpha2}) is reduced to
\begin{align}
&\dot{F}_{\pm }(\tau )+i(\omega _{0}-\lambda \kappa )F_{\pm }(\tau )
\notag
\\
&~~~~~~~+(1\mp \lambda )\int_{0}^{\tau }d\tau ^{\prime }\mu (\tau
-\tau ^{\prime })F_{\pm }(\tau ^{\prime })=0,  \label{ff}
\end{align}%
with $\mu (x)=\sum_{k}e^{-i\omega _{k}x}|g_{k}|^{2}$ and $\lambda
=\pm 1$. The explicit forms of $\Omega (t)$, $\Omega ^{\prime }(t)$,
and $\Gamma (t)$ in the master equation are given by
\begin{align}
\Omega (t) =&\omega _{0}+\mathop{\rm Im}\left[ G_{\lambda
}(t)\right] ,\text{ \ } \Omega ^{\prime }(t) = \kappa +\lambda
\mathop{\rm Im}\left[
G_{\lambda }(t)\right] ,  \notag \\
& \Gamma (t) =\lambda \Gamma ^{\prime }(t)=\mathop{\rm Re}\left[
G_{\lambda }(t)\right] ,  \label{parameters}
\end{align}
where
\begin{align}
G_{\lambda }(t) =&-\frac{1}{2}\text{\ }\Big[\frac{\dot{F}_{-\lambda
}(t)}{F_{-\lambda }(t)}+i(\omega _{0}+\lambda \kappa )\Big]  \notag \\
=&\frac{1}{F_{-\lambda }(t)}\int_{0}^{t}d\tau \mu (t-\tau
)F_{-\lambda }(\tau ).  \label{para2}
\end{align}%
This result has the similar form as the coefficients in the
non-Markovian master equation of a two-level atom derived in
\cite{Breuer02}.

To obtain the operator form of the master equation, we should
introduce the following functional differential relations in the
coherent-state representation (i.e., the Bargmann representation of
operators \cite {Vourdas94}),
\begin{align}
\bar{\alpha}_{l}\frac{\delta \rho
(\boldsymbol{\bar{\alpha}},\boldsymbol{\ \alpha }^{\prime
};t)}{\delta \bar{\alpha}_{m}} \longleftrightarrow
&a_{l}^{\dag }a_{m}\rho (t),  \notag \\
\frac{\delta \rho (\boldsymbol{\bar{\alpha}},\boldsymbol{\alpha
}^{\prime };t)}{\delta \alpha _{l}}\alpha _{m} \longleftrightarrow
&\rho
(t)a_{l}^{\dag }a_{m},  \notag \\
\frac{\delta ^{2}\rho (\boldsymbol{\bar{\alpha}},\boldsymbol{\alpha
} ^{\prime };t)}{\delta \bar{\alpha}_{l}\delta \alpha _{m}}
\longleftrightarrow &a_{l}\rho (t)a_{m}^{\dag },  \label{ssd}
\end{align}%
with which we arrive at an operator form of the master equation
shown below,
\begin{align}
\dot{\rho}(t) =&-\frac{i}{\hbar }[H^{\prime }(t),\rho (t)] \notag
\\
&+\Gamma'(t)\sum_{k\neq k'}[2a_{k}\rho (t)a_{k'}^{\dag }-a_{k}^{\dag
}a_{k'}\rho (t)-\rho (t)a_{k}^{\dag }a_{k'}] \notag
\\
&+\Gamma(t)\sum_{k=1} ^2 [2a_{k}\rho (t)a_{k}^{\dag }-a_{k}^{\dag
}a_{k}\rho (t)-\rho (t)a_{k}^{\dag }a_{k}], \label{mas}
\end{align}
where
\begin{equation}
H^{\prime }(t)=\hbar \Omega (t)(a_{1}^{\dag }a_{1}+a_{2}^{\dag
}a_{2})+\hbar \Omega ^{\prime }(t)(a_{1}^{\dag }a_{2}+a_{2}^{\dag
}a_{1}),
\end{equation}%
is the modified Hamiltonian of the two optical modes. From
Eq.~(\ref{mas}), we can see that besides the spontaneous decay of
the individual mode, the environment, even only the vacuum
fluctuation is considered, will result in a coherent interaction and
a correlated spontaneous decay between the two modes. More
importantly, our derivation of the master equation is fully
non-perturbative, which goes beyond the Born-Markov approximation
and contains all the back-actions between environment and the
optical modes. The non-Markovian character resides in the
time-dependent coefficients of the exact master equation. These
formulae have been used to study the non-Markovian entanglement
dynamics of two squeezed states \cite{An07}.

The time-dependent coefficients in the exact master equation,
determined by Eq.~(\ref{ff}), crucially depend on the so-called
spectral density, which characterizes the coupling strength of the
environment to the system with respect to the frequencies of the
environment. It is defined as $J(\omega )=\sum_{k}\left\vert
g_{k}\right\vert^{2}\delta (\omega -\omega _{l})$. In the continuum
limit the spectral density may have the form
\begin{equation}
J(\omega )=\eta \omega \Big( \frac{\omega }{\omega _{c}}\Big)^{n-1}
e^{- \frac{\omega }{\omega_{c}}} , \label{spectral}
\end{equation}
where $\omega _{c}$ is an exponential cutoff frequency, and $\eta $
is a dimensionless coupling constant. The environment is classified
as Ohmic if $ n=1$, sub-Ohmic if $0<n<1$, and super-Ohmic for $n>1$
\cite{Leggett87}. Different spectral densities manifest different
non-Markovian dynamics.

It is worth mentioning that an exact master equation has also been
obtained very recently for the system of two harmonic oscillators
bilinearly coupling with a thermal environment \cite{Chou07}, where
the master equation is derived in the Wigner representation rather
than the operator form of Eq.~(\ref{mas}). Also the bilinear
coupling in \cite{Chou07} is defined in terms of the coordinate
variables of harmonic oscillators which is different from the
coupling Hamiltonian we used in Eq.~(\ref{HM1}). In terms of quantum
optics language, the coupling between the system and the environment
used in \cite{Chou07} involves simultaneously photon-photon
scattering process and two-photon creation and annihilation process
with the same coupling strength. Note that photon-photon scatterings
are linear optical processes while two-photon creation and
annihilation processes are non-linear optical processes, they cannot
have the same coupling strength in quantum optics. Therefore, the
model used in \cite{Chou07} might describe a physical system quite
different from the optical system we considered in the present work.

\subsection{The Markovian approximation}

It is interesting to see that one can reproduce the conventional
Markov solution from our exact non-Markovian master equation under
certain approximation. By redefining the dynamical variables of the
system as $ \alpha _{l}(\tau )=x_{l}(\tau)e^{-i\omega _{0}\tau }$,
and $\bar{\alpha}_{l}^{\prime }(\tau )=x_{l}^{\prime
}(\tau)e^{i\omega _{0}\tau }$, we can recast
Eq.~(\ref{alpha1}-\ref{alpha2}) into
\begin{align}
\dot{x}_{l}+i\kappa x_{l^{\prime }}+\int_{0}^{\infty }d\omega
J(\omega )& \int_{0}^{\tau }d\tau ^{\prime }e^{i(\omega _{0}-\omega
)(\tau -\tau
^{\prime })}[x_{l}\left( \tau ^{\prime }\right)  \notag \\
&+\lambda x_{l^{\prime }}\left( \tau
^{\prime }\right) ]=0, \label{x1} \\
\dot{\bar{x}}_{l}^{\prime }-i\kappa \bar{x}_{l^{\prime }}^{\prime
}+\int_{0}^{\infty }d\omega J(\omega ) & \int_{0}^{\tau }d\tau
^{\prime }e^{-i(\omega _{0}-\omega )(\tau -\tau ^{\prime
})}[\bar{x}_{l}^{\prime
}\left( \tau ^{\prime }\right)  \notag \\
&+\lambda \bar{x}_{l^{\prime }}^{\prime }\left( \tau ^{\prime
}\right) ]=0.  \label{x2}
\end{align}
Then, we take the so-called Markov approximation,
\begin{equation}  \label{Mlimit}
x\left( \tau ^{\prime }\right) \cong x(\tau),\text{ \
}\bar{x}^{\prime }\left( \tau ^{\prime }\right) \cong
\bar{x}^{\prime }\left( \tau \right) ,
\end{equation}
namely, approximately taking the dynamical variables to the ones
that depend only on the present time so that any memory regarding
the earlier time is ignored \cite{Milonni}.

The Markov approximation is mainly based on the physical assumption
that the correlation time of environment is very small compared with
the typical time scale of system evolution. Also under this
assumption we can extend the upper limit of the $\tau ^{\prime }$
integration in Eqs.~(\ref{x1}-\ref{x2}) to infinity and use the
equality
\begin{align}
\lim_{\tau \rightarrow \infty }\int_{0}^{\tau }d\tau ^{\prime
}e^{\pm
i(\omega _{0}-\omega )(\tau -\tau ^{\prime })}= ~~~~~~~~~~~  \notag \\
\pi \delta (\omega -\omega _{0})\mp i\mathscr{P}\Big(\frac{1}{\omega
-\omega _{0}}\Big),  \label{iden}
\end{align}
where $\mathscr{P}$ and the delta-function denote the Cauchy
principal value and the singularity, respectively. The
integro-differential equations in (\ref{x1}-\ref{x2}) are thus
reduced to a couple of linear ordinary differential equations. The
solutions of $x_{l}$ and $\bar{x}_{l}^{\prime }$, as well as $
\alpha _{l}$ and $\bar{\alpha}_{l}^{\prime }$ can then be easily
obtained, which result in
\begin{align}
u =&\frac{e^{-i(\omega _{0}-\lambda \kappa )\tau }+e^{[-i(\omega
_{0}+\lambda \kappa )-2(\pi J(\omega _{0})-i\delta \omega )]\tau
}}{2},
\label{u} \\
v =&\frac{e^{-i(\omega _{0}-\lambda \kappa )\tau }-e^{[-i(\omega
_{0}+\lambda \kappa )-2(\pi J(\omega _{0})-i\delta \omega )]\tau
}}{2\lambda },  \label{v}
\end{align}%
where $\delta \omega =\mathscr{P}\int_{0}^{\infty }\frac{J(\omega
)d\omega }{ \omega -\omega _{0}}$. Using the solutions
(\ref{u}-\ref{v}), one can verify from Eqs. (\ref{go}) that,
\begin{align}
\Gamma (t) =&\lambda \Gamma ^{\prime }(t)=\pi J(\omega _{0}),  \notag \\
\Omega (t) =&\omega _{0}-\delta \omega ,\text{ \ }\Omega ^{\prime
}(t)=\kappa -\lambda \delta \omega ,  \label{Ml}
\end{align}%
which is exactly the coefficients in the Markov master equation of
the optical system \cite{Carmichael93}. This result can also be
obtained easier by directly applying the Markov approximation
Eqs.~(\ref{Mlimit}-\ref {iden}) to
Eqs.~(\ref{parameters}-\ref{para2}).

As shown above, all the coefficients in the master equation have
become time-independent, and the non-Markovian master equation
(\ref{mas}) is reduced to the Markov master equation under the
Markov approximation. This Markov approximation is valid to all
kinds of spectral densities, including Ohmic, super-Ohmic and
sub-Ohmic cases, while a different spectral density does produce the
frequency shift, $\delta \omega=\mathscr{P}\int_{0}^{\infty
}\frac{J(\omega )d\omega }{ \omega -\omega _{0}}$, and decay rate,
$\Gamma = \pi J(\omega_0)$, differently. As a result, we conclude
that our exact non-Markovian master equation can not only explore
more complicated situation where Markov approximation is
unreachable, but also examine different spectral densities between
the system and the environment even in the Markovian limit. This
actually provides a simple way to reveal the underlying mechanism of
quantum decoherence.

\section{Decoherence dynamics of entangled coherent states}

There are two different types of continuous variable entangled
states. One is the entangled squeezed states, and the other is the
entangled coherent states\cite{Sanders92}. We have used the exact
non-Markovian master equation derived in this paper to study the
non-Markovian entanglement dynamics of two squeezed states in a
recent published paper \cite{An07}. In this section, we will analyze
the decoherence properties of the entangled coherent states. The
decoherence dynamics of the two coherent modes is also fully
described by the master equation (\ref{mas}) with the non-Markovian
character residing in its time-dependent coefficients. The
time-dependent coefficients in the master equation are determined by
$F_{\pm}(t)$ as the solution of Eq.~(\ref{ff}) for a specific
environmental spectral density. In the present work, we will
consider the Ohmic spectral density, i.e., $n=1$ in Eq.
(\ref{spectral}), which is often the case for optical communication.

\begin{figure} [htbp]
\centerline{\epsfig{file=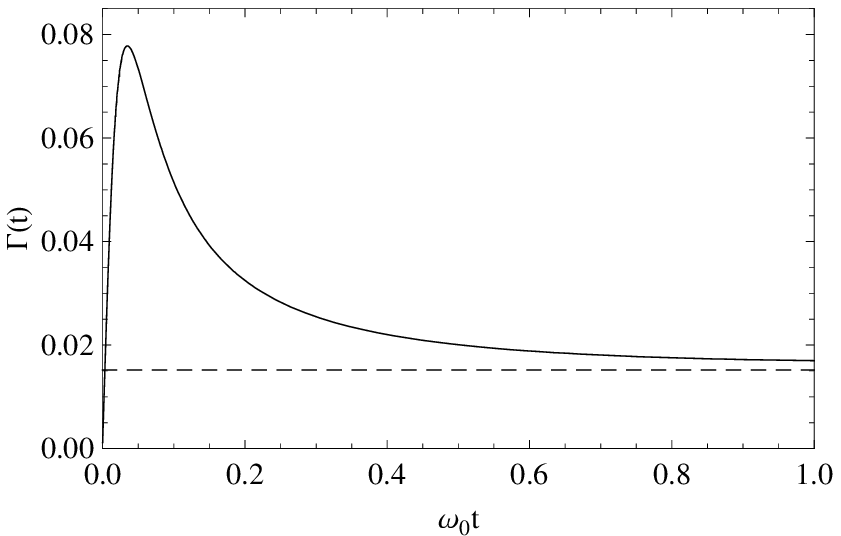, width=8.2cm}} 
\vspace*{13pt} \fcaption{\label{decay}Comparison of the decay rate
$\Gamma(t)$ [$=\lambda \Gamma^{\prime }(t)$] between the
non-Markovian (solid line) and Markovian (dashed line) results. The
parameters $\kappa/\omega_{0} =0.5 $,
$\protect\omega_{c}/\protect\omega_{0}=30.0$, and $\eta =0.005$ used
in the numerical calculation.}
\end{figure}

\begin{figure} [htbp]
\centerline{\epsfig{file=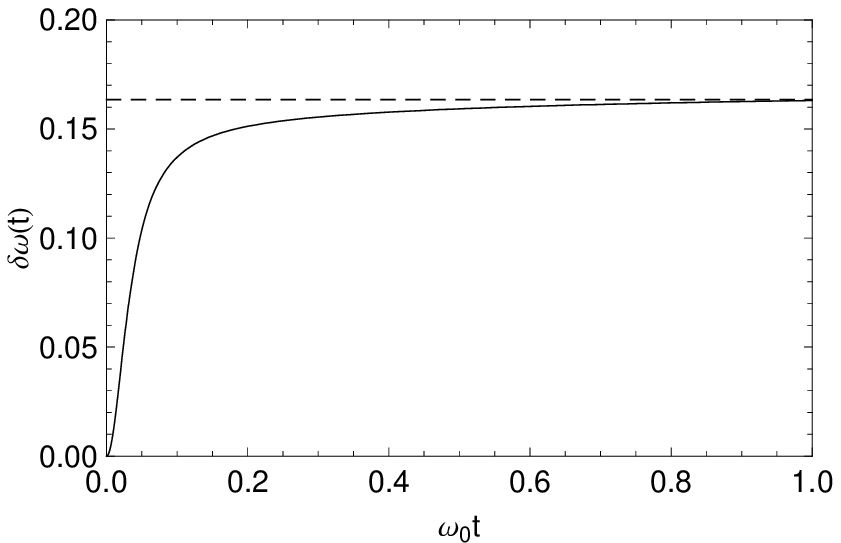, width=8.2cm}} 
\vspace*{13pt} \fcaption{\label{fre}Comparison of the frequency
shift $\protect\delta\protect\omega(t)$ between the non-Markovian
(solid line) and Markovian (dashed line) results. The parameters
used in the numerical calculation are the same as that in Fig.
\ref{decay}.}
\end{figure}

In Figs. \ref{decay} and \ref{fre}, we plot the numerical results of
the frequency shift $\delta \omega (t)$ and decay rate $\Gamma (t)$
of the individual optical field as well as their corresponding
Markovian values. It shows that the non-Markovian dissipation-noise
dynamics is characterized by two time scales: $\tau _{1}=1/\omega
_{c}$ (the shortest time scale of the environment) and $\tau
_{2}=1/\omega _{0}$ (the time scale of the optical modes). When
$t<\tau _{1}$, both coefficients, $\delta \omega (t)$ and $\Gamma
(t)$, grow very quickly, while after $\tau _{1}$, $\delta \omega
(t)$ and $\Gamma (t)$ approach to the corresponding Markov values,
given by Eq.~(\ref{Ml}), gradually when the time approaches to the
time scale $\tau _{2}$. It clearly evidences that the non-Markovian
effect has a huge deviation from the Markov effect within the time
scale $\tau _{2}$. This deviation will influence the dynamics later
on significantly as a historical memory effect. The time dependent
coefficients in the exact master equation (\ref{mas}) contain all
the back-action effects between the system and the environment. The
non-Markovian decoherence dynamics of the quantum optical field
system thus becomes transparent due to the sensitive time dependence
of these coefficients within the time scale $\tau _{2}$.

In the following, we will investigate the decoherence dynamics of
the entangled coherent states under the influence of the vacuum
fluctuation. The entangled coherent states are defined as
\begin{align}
|\psi _{\pm }\rangle =&\frac{1}{\sqrt{N_{\pm }}}\big(|\alpha
,-\alpha
\rangle \pm |-\alpha ,\alpha \rangle \big),  \notag \\
|\phi _{\pm }\rangle =&\frac{1}{\sqrt{N_{\pm }}}\big(|\alpha ,\alpha
\rangle \pm |-\alpha ,-\alpha \rangle \big),  \label{ecs}
\end{align}
which were studied as quasi-Bell states \cite{Sanders92,Wang02},
where $N_{\pm }=2(e^{2\left\vert \alpha \right\vert ^{2}}\pm
e^{-2\left\vert \alpha \right\vert ^{2}})$ are the normalization
constants. Many schemes to generate such states have been proposed
in optical systems and also in other systems
\cite{Agarwal03,Pater03,Armour02,Jeong06}. It has also proposed to
use these entangled coherent states for teleporting the superposed
coherent states \cite{van01,Zheng03}.

The time evolutions of these entangled coherent states are given by
\begin{eqnarray}
\rho _{\psi _{\pm }}(t) =&\frac{1}{N_{\pm }}\Big[e^{2(\left\vert
\alpha \right\vert ^{2}-\left\vert a_{+}(t)\right\vert
^{2})}\big(|a_{+}(t),-a_{+}(t)\rangle \langle
\bar{a}_{+}(t),-\bar{a}_{+}(t)|+|-a_{+}(t),a_{+}(t)\rangle \langle
-\bar{a}_{+}(t),\bar{a}_{+}(t)|\big)
\notag \\
&\pm e^{-2(\left\vert \alpha \right\vert ^{2}-\left\vert
a_{+}(t)\right\vert ^{2})}\big(|a_{+}(t),-a_{+}(t)\rangle \langle
-\bar{a} _{+}(t),\bar{a}_{+}(t)|+|-a_{+}(t),a_{+}(t)\rangle \langle
\bar{a}_{+}(t),-
\bar{a}_{+}(t)|\big)\Big],  \notag \\
\rho _{\phi _{\pm }}(t) =&\frac{1}{N_{\pm }}\Big[e^{2(\left\vert
\alpha \right\vert ^{2}-\left\vert a_{-}(t)\right\vert
^{2})}\big(|a_{-}(t),a_{-}(t)\rangle \langle \bar{a}_{-}(t),\bar{a}
_{-}(t)|+|-a_{-}(t),-a_{-}(t)\rangle \langle
-\bar{a}_{-}(t),-\bar{a}_{-}(t)|\big)  \notag \\
&\pm e^{-2(\left\vert \alpha \right\vert ^{2}-\left\vert
a_{-}(t)\right\vert ^{2})}\big(|a_{-}(t),a_{-}(t)\rangle \langle
-\bar{a} _{-}(t),-\bar{a}_{-}(t)|+|-a_{-}(t),-a_{-}(t)\rangle
\langle \bar{a}_{-}(t), \bar{a}_{-}(t)|\big)\Big],  \label{gw}
\end{eqnarray}%
respectively, where $a_{\pm }(t)=\alpha F_{\pm }(t)$. Eqs.~(
\ref{gw}) can be obtained directly from the exact solution of the
reduced density matrix, Eq.~(\ref{rout}) plus Eq.~(\ref{prord}) by
integrating over the initial variables. From Eq.~(\ref{ff}) one can
verify that for $\lambda =1$, the entangled coherent states $|\psi
_{\pm }\rangle $ remain in pure states (decoherence free states)
because $a_{+}(t)=\alpha F_{+}(t)=\alpha e^{-i(\omega _{0}-\kappa
)t}$, namely the time evolution of $|\psi _{\pm }\rangle $ is
independent of the decay rate $\Gamma (t)$ and the shift frequency
$\delta \omega (t)$ which are determined by $F_{-}(t)$ when $
\lambda =1$ [see Eqs.~(\ref{parameters}) and (\ref{para2})] and only
affect on the time evolution of the other two entangled coherent
state $|\phi _{\pm }\rangle $. Similarly, when $\lambda =-1$, the
entangled coherent states $ |\phi _{\pm }\rangle $ becomes
decoherence free states since $ a_{-}(t)=\alpha F_{-}(t)=\alpha
e^{-i(\omega _{0}+\kappa )t}$, while the decay rate $\Gamma (t)$ and
the shift frequency $\delta \omega (t)$ are determined by $F_{+}(t)$
which only affects the states $|\psi _{\pm }\rangle $. Since
$\lambda =1$ (or $-1$) corresponds to the case of the two optical
modes coupling to the environment in phase (or out of phase), the
above result indicates that two of the four entangled coherent
states in Eq.~(\ref {ecs}) become decoherence-free states
\cite{An05} if the two optical modes couple to the environment in
phase (a constructive interference coupling) or out of phase (a
destructive interference coupling).

The reason that the $\psi $-type and $\phi $-type entangled coherent
states in Eq.~(\ref{ecs}) have different decoherence behaviors comes
from different symmetric properties of these entangled coherent
states. The $\psi $-type and $\phi $-type coherent states correspond
to the center-of-mass and relative motions of two-field coherent
states, respectively. This property becomes clear by defining the
center-of-mass and relative motional variables of the two subsystems
as $A^{\dag }=(a_{1}^{\dag }+a_{2}^{\dag })$ and $ a^{\dag
}=(a_{1}^{\dag }-a_{2}^{\dag })$. As one can find, $|\psi _{\pm
}\rangle $ consist of only the relative motion, while $|\phi _{\pm
}\rangle $ lie only on the center-of-mass motion. When the two
optical modes couple to the environment in phase, namely,
$g_{1k}=g_{2k}=g_{k}$, the interaction between the optical modes and
the environment only affects the center-of-mass motion so that the
entangled coherent states of the relative motion, $|\psi _{\pm
}\rangle $, become decoherence-free states. On the other hand, if
the two optical modes couple to the environment out of phase, i.e.
$g_{1k}=-g_{2k}=g_{k}$, the interaction between them only affects
the relative motion but leaves the entangled coherent states of the
center-of-mass motion, $|\phi _{\pm }\rangle $, free from
decoherence. This is indeed a consequence of the sufficient
condition for the decoherence-free space protected by
symmetry\cite{Lidar}.

We shall quantify the entanglement degree of the entangled coherent
states by the familiar concept of concurrence usually used in a
discrete basis \cite{Wootters98}. To do so, we may rewrite
Eq.~(\ref{gw}) in terms of the orthogonal basis \cite{Wang02},
\begin{align}
|\boldsymbol{0}\rangle =&e^{-\frac{\left\vert a_{\pm }(t)\right\vert
^{2}}{2 }}|a_{\pm }(t)\rangle ,  \notag \\
|\boldsymbol{1}\rangle =&\frac{e^{-\frac{\left\vert a_{\pm
}(t)\right\vert ^{2}}{2}}|-a_{\pm }(t)\rangle -p_{\pm
}(t)|\boldsymbol{0}\rangle }{\sqrt{ 1-p_{\pm }(t)^{2}}},
\end{align}
with $p_{\pm }(t)=e^{-2\left\vert a_{\pm }(t)\right\vert ^{2}}$.
Above change from the coherent state basis to the
$|\boldsymbol{0}\rangle$ and $| \boldsymbol{1}\rangle$ basis is
equivalent to a local unitary transformation of the states, which
does not modify the entanglement degree in the original states. In
this discrete basis, the concurrence can be calculated as usual. It
is not difficult to find that the concurrence $C_{\phi_{-}}(0)=C_{
\psi_{-}}(0)=1$, which is maximally entangled irrespective of the
amplitude $ \alpha$, while $C_{\phi_{+}}(0)=C_{\psi_{+}}(0)=\tanh
2\left\vert \alpha \right\vert^{2}$, which imply that the $\phi_+$
and $\psi_+$ states are not initially maximally entangled. One can
also show that when the two optical modes couple with the
environment in phase, i.e. $\lambda=1$, the concurrence $C_{\psi
_{-}}(t)=1$, and $C_{\psi _{+}}(t)=\tanh 2\left\vert \alpha
\right\vert^{2}$, namely, the entanglement of $|\psi_{\pm} \rangle$
remain unchanged during the time evolution. While
$|\phi_{\pm}\rangle$ are sensitive to decoherence. In contrast, if
the two optical modes interact with the environment out of phase,
i.e. $\lambda=-1$, $C_{\phi _{-}}(t)=1$, and $C_{\phi _{+}}(t)=\tanh
2\left\vert \alpha \right\vert^{2}$, while $ C_{\psi_{-}}(t)$ and
$C_{\psi _{+}}(t)$ will decay (disentanglement) due to the
decoherence.

\begin{figure} [htbp]
\centerline{\epsfig{file=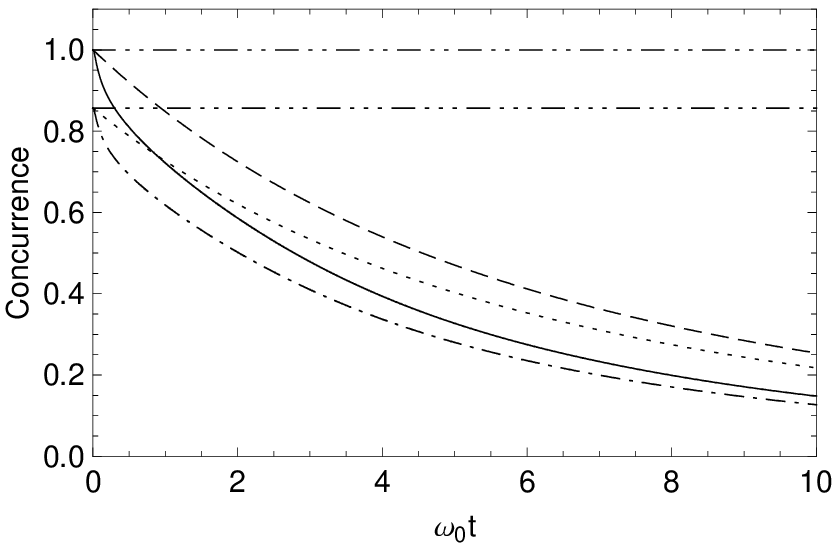, width=8.2cm}} 
\vspace*{13pt} \fcaption{\label{concu}Time evolution of the
concurrences for different initial states. The solid and dot-dashed
lines show the non-Markovian time evolution of the concurrences for
$|\phi _{-}\rangle$ and $|\phi _{+}\rangle$ with $\protect \lambda
=1$ (or $|\psi_{-}\rangle$ and $|\psi _{+}\rangle$ with $\lambda
=-1$), respectively. Their corresponding Markovian behaviors are
shown as the dashed and dotted lines, respectively. The
dot-dot-dashed and dot-dot-dot-dashed lines are the concurrences for
$|\psi _{-}\rangle$ and $|\psi _{+}\rangle$ with $\lambda =1$ (or
$|\phi _{-}\rangle$ and $|\phi _{+}\rangle$ with $\protect\lambda
=-1$), respectively, which remain unchanged during the time
evolution. The initial coherent state parameter $ \alpha =0.8$, and
other parameters are the same as in Fig. \ref{decay}.}
\end{figure}

In Fig.~3, we show the concurrence evolution in time for the
entangled coherent states $|\phi _{\pm}\rangle$ and $|\psi
_{\pm}\rangle$. With the in-phase coupling between the optical modes
and the environment ($ \lambda =1$), our numerical results verify
that the entanglement degrees of $ |\phi _{\pm}\rangle$ (given by
the solid and dot-dashed lines in Fig.~\ref{concu}) suffer from a
fast decay during the time evolution while the entanglement degrees
of $|\psi _{\pm}\rangle$ remain unchanged (the dot-dot-dashed and
dot-dot-dot-dashed lines in Fig.~\ref{concu}). To compare the
non-Markovian entanglement dynamics with the Markovian dynamics, we
also plot the concurrence evolution for $|\phi_{\pm}\rangle$ under
the Markovian approximation, denoted by the dashed and dotted lines,
respectively, in Fig.~\ref{concu}. As one can see, the non-Markovian
effect accelerates the disentanglement. This is mainly a
contribution of the short time peak in the decay rate $ \Gamma (t)$
as a memory effect. It is also worth noting that no entanglement
oscillator is observed in above solution even there has coherent
coupling $ \Omega ^{\prime }(t)$ presented. This is because the
two-optical-field coupling $\Omega ^{\prime }(t)$ contributes only a
global phase to the entangled coherent states during the time
evolution, which has no influence on the entanglement degree of the
states. While as expected, $|\psi _{\pm }\rangle$ are
decoherence-free irrespective of Markovian or non-Markovian dynamics
being considered.

For the case of out-of-phase coupling between the optical modes and
the environment, namely $\lambda=-1$, the roles of the decoherence
effect on the $\phi$-type and the $\psi$-type entangled coherent
states are exchanged. The $\phi$-type entangled coherent states
remain unchanged, while the $\psi$ -type entangled coherent states
are disentangled by decoherence. The numerical results of the
entanglement evolution for $|\psi_{\pm}\rangle$ and
$|\phi_{\pm}\rangle$ are given by the same curves in Fig.
\ref{concu} with the exchange between $|\psi_{\pm}\rangle$ and
$|\phi_{\pm}\rangle$ states as in the in-phase coupling case.

From the above analysis, one can find that when the two identical
optical modes couple to a common environment in an arbitrary phase
difference, no decoherence-free entangled coherent state can exist
among the four entangled coherent states. All the four entangled
states could be disentangled by the vacuum fluctuation in time. The
non-Markovian dynamics will speed up the disentanglement process
with respect to the Markovian approximation. However, the parameter
$\lambda $ models the phase difference between the two optical modes
coupling with the environment. Physically, it is always possible to
adjust the two optical modes such that the couplings of the two
optical fields with the environment are either in phase
($\lambda=-1$) or out of phase ($\lambda=-1$). Then two decoherence
free states among the four entangled coherent states can always be
constructed in principle. It is certainly interested in seeing
experimental evidences on the preservation of two decoherence free
entangled coherent states as well as the non-Markovian
disentanglement enhancement to the other two entangled coherent
states.

\section{Summary and Discussions}

In summary, we have studied the detrimental effects of environment
on the entangled coherent states. We microscopically modeled the
decoherence dynamics of entangled coherent states under the
influence of vacuum fluctuation. An exact master equation with
time-dependent coefficients reflecting the full memory effect of the
reduced system has been derived by using the Feynman-Vernon
influence functional theory in the coherent-state path-integral
representation, which enables us to treat both of the back-actions
from the environment to the system and from the system to the
environment self-consistently. In addition, we have also explicitly
deduced to the well-known Markovian dynamics for the optical modes
from our exact non-Markovian master equation in the Markovian
approximation. The analytical analysis of the difference between the
non-Markovian dynamics and its Markovian approximation presented in
this paper may provide a quantitative way to experimentally explore
the non-Markovian effect as well as the spectral densities between
the system and the environment even in the Markovian limit.

We then investigated the non-Markovian dynamics of the entangled
coherent states, one of two typical continuous variable entanglement
states often used in quantum information processing. The other type
of continuous variable squeezed states has already be studied by two
of us based the same master equation derived here \cite{An07}. Our
first-principle analysis shows that the non-Markovian effect
accelerates the disentanglement compared with the results based on
Markovian approximation. It is the short time peak of the time
dependent coefficients in the master equation, which is incorporated
with the system's dynamics as a historical memory effect, that
contributed to this acceleration. Although the Born-Markovian
approximation has been widely employed in the field of quantum
optics, we argue that our investigation might be helpful for
understanding decoherence in nanoscale cavity devices and ultrafast
optical processes. For example, we have noticed the rapid
development of optical cavity technology, which has been employed to
confine a single atom \cite{kimble} or a single quantum dot
\cite{Altug,imma} in strong coupling regime, the prerequisite of
quantum network. The strong interaction occuring in nanometer size
subject to vacuum fluctuation suffers from some unpredictable
incoherence errors \cite{kimble}, which should involve the
non-Markovian effects. In this sense, our study of non-Markovian
dynamics, although with a simplified model, paves a way toward
clarification of the mechanism regarding those incoherence sources.

We have also shown how the decoherence behaviors of the different
entangled coherent states depend on the symmetrical properties of
these entangled coherent states as well as the interference
properties of couplings between the two optical modes with the
vacuum electromagnetic environment. Since the exact non-Markovian
master equation has been derived non-perturbatively and exactly,
decoherence dynamics subject to different spectral densities of
environment would be naturally available by our treatment. In fact,
the non-Markovian master equation (i.e., Eq.~(\ref{mas})) derived in
this paper has been used in treating the decoherence dynamics of
entanged squeezed states with sub-Ohmic and super-Ohmic spectral
densities of the environment \cite{An07}. More complicated cases,
e.g., the environment at finite temperature, would be hopefully
figured out by the similar way to the derivation of
Eq.~(\ref{prord}). As a final remark, we would like to mention a
very recent experiment for distinguishing different coherent states
\cite{cook}, which shows a potential of using entangled coherent
states for quantum communication. The entangled coherent states have
significantly different properties from the entangled squeezed
states and have been proposed as another type of continuous variable
quantum channels. As robustness of the quantum channel is essential
in view of decoherence, we expect that our consideration of
decoherence dynamics of entangled coherent states would be useful
for understanding quantum communication experiments with realistic
ultrafast optical processes in nanocavities and nanophotonic
systems.

\section*{Acknowledgement}

We would like to thank B. L. Hu, M. T. Lee and W. Y. Tu for useful
discussions. This work is supported by the National Science Council
of ROC under Contract No.~NSC-95-2112-M-006-001,
No.~NSC-94-2120-M-006-003, No.~NSC-96-2119-M-006-001, and by NNSF of
China under Grants No. 10604025 and No. 10774163, and Lzu05-02.


\begin{thebibliography}{99}
\bibitem{Bouwmeester97} D. Bouwmeester, J.-W. Pan, K. Matter, M. Eibl, H.
Weinfurter, and A. Zeilinger (1997), \textit{Experimental quantum
teleportation}, Nature, 390, pp. 575.

\bibitem{Braunstein98} S. L. Braunstein and H. J. Kimble (1998),
\textit{Teleportation of continuous quantum variables}, Phys. Rev.
Lett., 80, pp. 869.

\bibitem{Furusawa98} A. Furusawa, J. L. Sorensen, S. L. Braunstein, C. A.
Fuchs, H. J. Kimble, and E. S. Polzik (1998), \textit{Unconditional
quantum teleportation}, Science, 282, pp. 706.

\bibitem{Wolf07} M. M. Wolf, D. P\'{e}rez-Garc\'{\i}a, and G. Giedke (2007),
\textit{Quantum capacities of bosonic channels}, Phys. Rev. Lett.,
98, pp. 130501.

\bibitem{Sanders92} B. C. Sanders (1992), \textit{Entangled coherent states},
Phys. Rev. A, 45, pp. 6811.

\bibitem{Gerry97} Christopher C. Gerry (1997), Phys. Rev. A \textbf{55}, 2478.

\bibitem{Agarwal03} E. Solano, G. S. Agarwal and H. Walther (2003),
\textit{Strong-driving-assisted multipartite entanglement in cavity
QED}, Phys. Rev. Lett., 90, pp. 027903.

\bibitem{Pater03} M. Paternostro and M. D. Kim (2003), \textit{Generation
of entangled coherent states via cross-phase-modulation in a double
electromagnetically induced transparency regime}, Phys. Rev. A, 67,
pp. 023811.

\bibitem{Armour02} A. D. Armour, M. P. Blencowe and K. C. Schwab (2002),
 \textit{Entanglement and decoherence of
a micromechanical resonator via coupling to a Cooper-Pair box},
Phys. Rev. Lett., 88, pp. 148301.

\bibitem{Bose06} S. Bose and G. S. Agarwal (2006), \textit{Entangling pairs
of nano-cantilevers, Cooper-pair boxes and mesoscopic
teleportation}, New J. Phys., 8, pp. 34.

\bibitem{van01} S. J. van Enk and O. Hirota (2001), \textit{Entangled
coherent states: Teleportation and decoherence}, Phys. Rev. A, 64,
pp. 022313.

\bibitem{Wang01} X. Wang (2001), \textit{Quantum teleportation of entangled
coherent states}, Phys. Rev. A, 64, pp. 022302.

\bibitem{Jeong01} H. Jeong, M. S. Kim, and J. Lee (2001), \textit{Quantum-information
processing for a coherent superposition state via a mixedentangled
coherent channel}, Phys. Rev. A, 64, pp. 052308.

\bibitem{Jakub04} J. S. Prauzner-Bechcicki (2004), \textit{Two-mode squeezed vacuum
state coupled to the common thermal reservoir}, J. Phys. A: Math.
Gen., 37, pp. L173.

\bibitem{An05} J.-H. An, S.-J. Wang, and H.-G. Luo (2005), \textit{Entanglement
production and decoherence-free subspace of two single-mode cavities
embedded in a common environment}, J. Phys. A: Math. Gen., 38, pp.
3579.

\bibitem{Rossi06} R. Rossi Jr., A. R. Bosco de Magalh\~{a}es, M. C. Nemes
(2006), \textit{Two cavity modes in a dissipative environment: Cross
decay rates and robust states}, Physica A, 365, pp. 402.

\bibitem{Adesso} G. Adesso, A. Serafini, and F. Illuminati (2006), \textit{Multipartite
entanglement in three-mode Gaussian states of continuous-variable
systems: Quantification, sharing structure, and decoherence}, Phys.
Rev. A, 73, pp. 032345.

\bibitem{Ban06} M. Ban (2006), \textit{Decoherence of continuous variable quantum
information in non-Markovian channels}, J. Phy. A: Math. Gen., 39,
pp. 1927.

\bibitem{Goan07} K. L. Liu and H. S. Goan (2007), \textit{Non-Markovian entanglement
dynamics of quantum continuous variable systems in thermal
environments}, Phys. Rev. A, 76, pp. 022312.

\bibitem{Redfield65} A. G. Redfield (1965), \textit{The Theory of Relaxation Processes},
 Adv. Magn. Reson., 1, pp. 1.

\bibitem{Lindblad76} G. Lindblad (1976), \textit{On the generators of quantum
dynamical semigroups}, Commun. Math. Phys., 48, pp. 119.

\bibitem{Carmichael93} H. J. Carmichael (1993), \textit{An Open Systems Approach to
Quantum Optics}, Lecture Notes in Physics, Vol. m18
(Springer-Verlag, Berlin).

\bibitem{blatt07} F. Dublin, D. Rotter, M. Mukherjee, C. Russo, J. Eschner,
and R. Blatt (2007), \textit{Photon correlation versus interference
of single-Atom fluorescence in a half-cavity}, Phys. Rev. Lett., 98,
pp. 183003.

\bibitem{John94} S. John and T. Quang (1994), \textit{Localization of superradiance
near a photonic band gap}, Phys. Rev. Lett., 74, pp. 3419.

\bibitem{Aquino04} G. Aquino, L. Palatella, and P. Grigolini (2004), \textit{Absorption
and emission in the non-Poissonian case}, Phys. Rev. Lett., 93, pp.
050601.

\bibitem{Budini06} A. A. Budini (2006), \textit{Non-Poissonian intermittent fluorescence
from complex structured environments}, Phys. Rev. A, 73, pp.
061802(R).

\bibitem{mtlee06} M. T. Lee and M. W. Zhang (2006), \textit{Decoherence induced by
electron accumulation in a quantum measurement of charge qubits},
Phys. Rev. B, 74, pp. 085325.

\bibitem{Falci05} G. Falci, A. D'Arrigo, A. Mastellone, and E. Paladino
(2005), \textit{Initial decoherence in solid state qubits}, Phys.
Rev. Lett., 94, pp. 167002.

\bibitem{Breuer02} H.-P. Breuer and F. Petruccione (2002), \textit{The theory of
open quantum systems}, (Oxford University Press, Oxford).

\bibitem{Maniscalco07} S. Maniscalco, S. Olivares, and M. G. A. Paris (2007),
\textit{Entanglement oscillations in non-Markovian quantum
channels}, Phys. Rev. A, 75, pp. 062119.

\bibitem{Chou07} C. H. Chou, T. Yu, and B. L. Hu (2008), \textit{Exact Master
Equation and Quantum Decoherence of Two Coupled Harmonic Oscillators
in a General Environment}, Phys. Rev. E, 77, pp. 011112.

\bibitem{An07} J.-H. An, and W. M. Zhang (2007), \textit{Non-Markovian entanglement
dynamics of noisy continuous-variable quantum channels}, Phys. Rev.
A, 76, pp. 042127.

\bibitem{Feynman63} R. P. Feynman and F. L. Vernon (1963), \textit{The theory of
a general quantum system interacting with a linear dissipative
system}, Ann. Phys. (N.Y.), 24, pp. 118.

\bibitem{Caldeira83} A. O. Caldeira and A. J. Leggett (1983), \textit{Path integral
approach to quantum Brownian motion}, Physic A, 121, pp. 587.

\bibitem{Hu92} B. L. Hu, J. P. Paz, and Y. Zhang (1992), \textit{Quantum Brownian
motion in a general environment: Exact master equation with nonlocal
dissipation and colored noise}, Phys. Rev. D, 45, pp. 2843.

\bibitem{wzhang90} W. M. Zhang, D. H. Feng, and R. Gilmore (1990), \textit{Coherent
states: Theory and some applications}, Rev. Mod. Phys., 62, pp. 867.

\bibitem{Wootters98} W. K. Wootters (1998), \textit{Entanglement of formation of
an arbitrary state of two qubits}, Phys. Rev. Lett., 80, pp. 2245.

\bibitem{Werner02} G. Vidal and R. F. Werner (2002), \textit{Computable measure
of entanglement}, Phys. Rev. A, 65, pp. 032314.

\bibitem{pelli} T. Pellizzari (1997), \textit{Quantum Networking with Optical
Fibres}, Phys. Rev. Lett., 79, pp. 5242.

\bibitem{li} Z. Q. Yin and F. L. Li (2007), \textit{Multiatom and resonant
interaction scheme for quantum state transfer and logical gates
between two remote cavities via an optical fiber}, Phys. Rev. A, 75,
pp. 012324.

\bibitem{Plenio06} M. J. Hartmann, F. G. S. L. Brand\~{a}o, and M. B.
Plenio (2006), \textit{Strongly interacting polaritons in coupled
arrays of cavities}, Nature Physics, 2, pp. 849.

\bibitem{Hollenberg06} A. D. Greentree, C. Tahan, J. H. Cole, L. C. L.
Hollenberg (2006), \textit{Quantum phase transitions of light},
Nature Physics, 2, pp. 856.

\bibitem{Angelakis08} D. G. Angelakis and A. Kay (2008), \textit{Cluster state
quantum computation in coupled cavity arrays }, New J. Phys., 10,
pp. 023012.

\bibitem{Armani03} D. K. Armani, T. J. Kippenberg, S. M. Spillane, and K. J.
Vahala (2003), \textit{Ultra-high-Q toroid microcavity on a chip},
Nature, 421, pp. 925.

\bibitem{Song05} B.-S. Song, S. Noda, T. Asano, and Y. Akahane (2005),
\textit{Ultra-high-Q photonic double-heterostructure nanocavity},
Nature Materials, 4, pp. 207.

\bibitem{Aoki06} T. Aoki, B. Dayan, E. Wilcut, W. P. Bowen, A. S. Parkins,
T. J. Kippenberg, K. J. Vahala, and H. J. Kimble (2006),
\textit{Observation of strong coupling between one atom and a
monolithic microresonator}, Nature, 443, pp. 671.

\bibitem{Anastopoulos} C. Anastopoulos and B.L. Hu (2000), \textit{Two-level
atom-field interaction: Exact master equations for non-Markovian
dynamics, decoherence, and relaxation}, Phys. Rev. A, 62, pp.
033821.

\bibitem{Leggett87} A. J. Leggett, S. Chakravarty, A. T. Dorsey, M. P. A.
Fisher, A. Garg, and W. Zwerger (1987), \textit{Dynamics of the
dissipative two-state system}, Rev. Mod. Phys., 59, pp. 1.

\bibitem{Klauder79} J. R. Klauder (1979), \textit{Path integrals and
stationary-phase approximations}, Phys. Rev. D, 19, pp. 2349.

\bibitem{Yu} T. Yu and J. H. Eberly (2004), \textit{Finite-time disentanglement
via spontaneous emission}, Phys. Rev. Lett., 93, pp. 140404; B.
Bellomo, R. Lo Franco, and G. Compagno (2007), \textit{Non-Markovian
Effects on the Dynamics of Entanglement}, Phys. Rev. Lett., 99, pp.
160502 .

\bibitem{Vourdas94} A. Vourdas and R.F. Bishop (1994), \textit{Thermal coherent
states in the Bargmann representation}, Phy. Rev. A, 50, pp. 3331.

\bibitem{Milonni} P. W. Milonni (1994), \textit{The Quantum Vacuum: An Introduction
to Quantum Electrodynamics}, (Academic, New York).

\bibitem{Wang02} X. Wang and B. C. Sanders (2001), \textit{Multipartite entangled
coherent states}, Phys. Rev. A, 65, pp. 012303.

\bibitem{Jeong06} H. Jeong and N. B. An (2006), \textit{Greenberger-Horne-Zeilinger
type and W-type entangled coherent states: Generation and Bell-type
inequality tests without photon counting}, Phys. Rev. A, 74, pp.
022104.

\bibitem{Zheng03} Y.-Z. Zheng, Y.-J. Gu, and G.-C. Guo (2003), \textit{Teleportation
of a coherent superposition state via a nonmaximally entangled
coherent channel}, J. Opt. B: Quantum Semiclass. Opt., 5, pp. 29.

\bibitem{Lidar} D. A. Lidar, I. L. Chuang and K. B. Whaley (1998),
\textit{Decoherence-free subspaces for quantum computation}, Phys.
Rev. Lett., 81, pp. 2594.

\bibitem{kimble} A. D. Boozer, A. Boca, R. Miller, T. E. Northup, and H. J.
Kimble (2007), \textit{Reversible State Transfer between Light and a
Single Trapped Atom}, Phys. Rev. Lett., 98, pp. 193601.

\bibitem{Altug} H. Altug, D. Englund, and J. Vuckovic (2006), \textit{Ultrafast
photonic crystal nanocavity laser}, Nature Physics, 2, pp. 484.

\bibitem{imma} K. Hennessy, A. Badolato, M. Winger, D. Gerace, M. Atature,
S. Gulde, S. Falt, E. L. Hu, A. Imamoglu (2007), \textit{Quantum
nature of a strongly coupled single quantum dot-cavity system},
Nature, 445, pp. 896.

\bibitem{cook} R. L. Cook, P. J. Martin, and J. M. Geremia (2007),
\textit{Optical coherent state discrimination using a closed-loop
quantum measurement}, Nature, 446, pp. 774.
\end{thebibliography}
\end{document}